\documentclass[amsmath,amssymb,aps,prb,
preprint,
floatfix]{revtex4-2}

\usepackage[utf8]{inputenc}
\usepackage{xcite}
\usepackage{xr}
\usepackage{amsmath}
\usepackage{graphicx}
\usepackage[hidelinks]{hyperref}
\usepackage[margin=1in]{geometry}

\usepackage{orcidlink}
\usepackage{xcolor}

\begin{document}
\title{Symmetry considerations in chirality-induced spin selectivity}

\author{Dmitry Budker\,\orcidlink{0000-0002-7356-4814}} 
\email{budker@uni-mainz.de}
\affiliation{Helmholtz-Institut Mainz, 55128 Mainz, Germany}
\affiliation{GSI Helmholtzzentrum f{\"u}r Schwerionenforschung GmbH, 64291 Darmstadt, Germany}
\affiliation{QUANTUM, Institut für Physik, Johannes Gutenberg-Universit{\"a}t, 55128, Mainz, Germany}
\affiliation{Department of Physics, University of California, 94720-7300, Berkeley, USA}
\author{Angela Wittmann\,\orcidlink{0000-0002-0005-8048}}
\email{a.wittmann@uni-mainz.de}
\affiliation{Institute of Physics, Johannes Gutenberg-University Mainz, 55128 Mainz, Germany}

\begin{abstract}
The chirality-induced spin selectivity (CISS) effect, the coupling between structural chirality and electron spin polarization, has been experimentally observed across many diverse systems. However, despite extensive theoretical effort, a unified mechanistic understanding remains elusive. In this perspective, we demonstrate how some of the basic properties of the fascinating effects of CISS can be understood based on straightforward symmetry considerations commonly employed in fundamental particle physics. In particular, we show that CISS does not violate any fundamental symmetries including parity and time-reversal. By anchoring CISS within the universal language of symmetry, we offer a robust conceptual foundation for interpreting experiments and guiding future theoretical and experimental designs.
\end{abstract}

\maketitle
\vspace{2cm}
\begin{center}
Dedicated to the memory of Prof. David Waldeck (1956-2026) and Prof. Vladimiro Mujica (1956-2026).
\end{center}
\newpage

\section{Introduction}

In summer of 2025, some 160 scientists from around the world, including Professors David Waldeck and Vladimiro Mujica, gathered for a week-long conference in New England to discuss the mechanisms and applications of the chirality-induced spin-selectivity (CISS) effect that has already left a profound mark on a remarkably diverse array of disciplines ranging from magnetism of layered magnetic materials to biology to magnetic resonance, and even geophysics.
Controlling electron spin, a key quantum mechanical property, using a macroscopic tool in the absence of magnetic field has been a longstanding challenge. The chirality-induced spin-selectivity effect offers a pathway to generating and controlling spin polarization via structural chirality. 

The CISS effect, first observed experimentally in 1999~\cite{Ray1999} and further explored in a series of pioneering papers~\cite{Hamelbeck2011,Xie2011,Mishra2013,BenDor2017,Banerjee-Ghosh2020,Nakajima2023,Eckvahl2023} is, in fact, a family of fascinating phenomena that connect molecular chirality with electron spins.
As with many things in science, one can trace earlier work anticipating CISS. For example, Ref.\,\cite{Pospelov1996} discussed the difference in scattering cross-sections on chiral molecules of electrons with opposite helicity, an effect similar to that of optical circular dichroism. The discussion was later extended to the effects in the interaction of unpolarized electrons and radiation with chiral molecules in the presence of magnetic field~\cite{Pospelov1996a}. 
From the early days of CISS to today, scientists have vigorously debated the origin of the effect and, for a long time, theoretical models were unable to predict the CISS effects quantitatively~\cite{Evers2022}.

\begin{figure}
\centering
\includegraphics[width=0.95\linewidth]{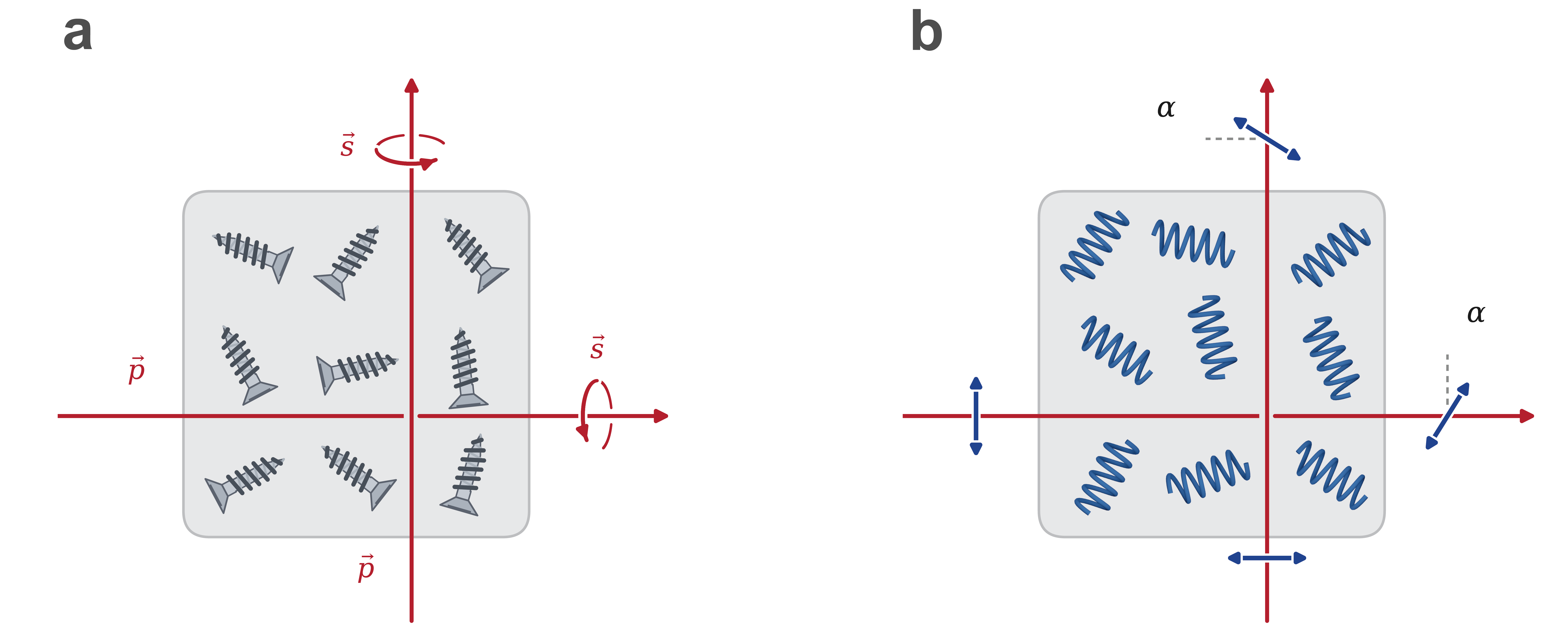}
\caption{Chirality as a pseudoscalar. (a) A medium of randomly oriented right-handed screws has no preferred direction and is invariant under time reversal, yet it is chiral. Probe balls sent along two perpendicular directions each emerge rotating, with the sense of the rotation $\vec{s}$ relative to the ball's own momentum $\vec{p}$ the same in both cases, so that $\langle\,\vec{s}\cdot\vec{p}\,\rangle\neq0$ irrespective of the direction of travel. (b) Replacing the screws with chiral molecules and the balls with beams of light turns the pseudoscalar $\chi$ into a source of optical rotation: along any path of light, the plane of polarization is rotated in the same sense, and by the same angle $\alpha$ for equal path lengths through the medium.}
\label{fig:chirality}
\end{figure}

Of central importance to our discussion here is the concept of chirality. In simple terms, chirality is a pseudoscalar property $(\chi)$ characterizing a physical system that changes sign upon spatial inversion (P) and remains invariant under time reversal (T)~\cite{Barron1986}. An example of such a system would be a collection of randomly oriented screws (see Fig.\,\ref{fig:chirality}a). Note that such a system does not have any preferred direction, is T-invariant and obviously chiral: it would rotate a probe particle (such as a wooden ball) propagating through the medium in a particular sense, clockwise in the case of right-handed screws. Replace the screws with chiral molecules and the wooden ball with a beam of light and you get optical rotation---the celebrated property of chiral solutions (Fig.\,\ref{fig:chirality}b).

While chirality is not necessarily associated with spin (as in the examples above), there is a plethora of chiral properties that can be associated with a spin or an ensemble of spins. This is discussed in detail in Ref.\,\cite{Kusunose_2024_Symmetries}, which provides a useful classification of the properties based on their various symmetries and tensor rank, along with examples of practical situation where such properties arise. For example, a single stationary spin $\vec{s}$ does not possess chirality; however, a moving spin can, indeed, be chiral via the \textit{correlation} also known as \textit{rotational invariant} $\vec{s}\cdot\vec{p}$, where $\vec{p}$ is the linear momentum of the particle with spin. 
Note that the $\vec{s}\cdot\vec{p}$ correlation can also be considered when the spin refers to one particle and the momentum to another, as we discuss below.

To study the fundamental symmetry properties of the different manifestations of the CISS effect, we subdivide the effects into \textit{dynamic CISS} and \textit{quasi-static CISS} configurations. Here, dynamic CISS refers to the generation of spin polarization when a charge current flows in the chiral system. This includes, for instance, experiments on photoemission from chiral substrates, magneto-transport experiments, where the conductivity of a hybrid magnetic-chiral heterostructure depends on the relative orientation of the current and magnetization, and charge transfer in isolated donor-chiral bridge-acceptor molecules \cite{Ray1999,Hamelbeck2011,Xie2011,Aragones2017,Inui2020,Safari2023,Eckvahl2023,Adhikari2023} as well as the Onsager reciprocal process, the inverse CISS effect~\cite{Lu2019a,Inui2020,Dong2025}.

The second category designates quasi-static CISS effects in the absence of an externally applied electrical bias at hybrid chiral molecule-metal interfaces. Prominent reports of this effect in the literature include enantioselective adsorption of chiral molecules on magnetic thin films~\cite{Banerjee-Ghosh2018}, magnetism induced by the proximity of chiral molecules (MIPAC)~\cite{BenDor2017}, and chirality-induced anisotropy in spin injection and spin-charge conversion~\cite{Sun2024a,Moharana2025} with wide-ranging implications for enantiospecific chemistry and the origin of homochirality of life.

\section{Symmetry properties}
\subsection{Time reversal and parity}
\begin{figure}
\centering
\includegraphics[width=0.55\linewidth]{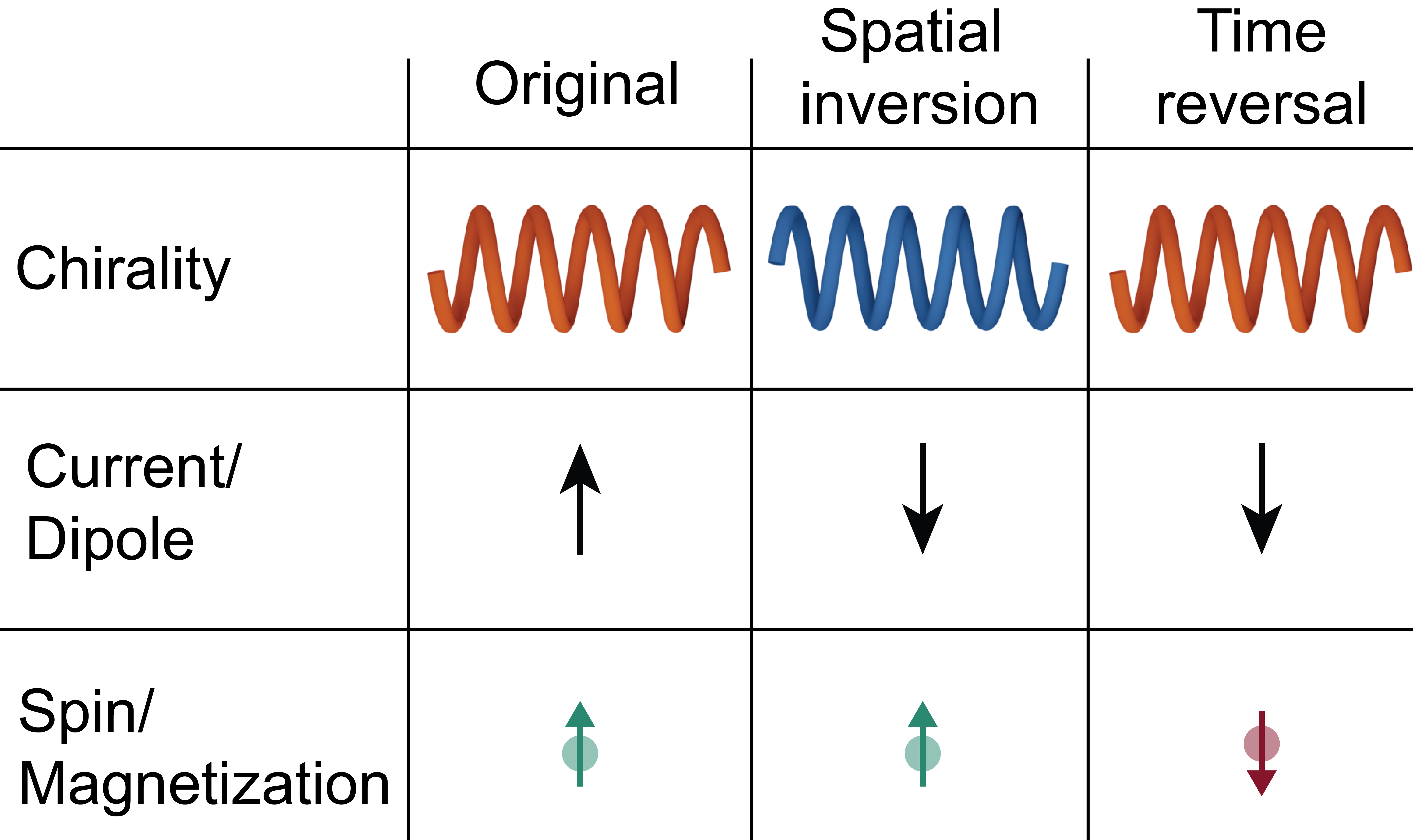}
\caption{Behavior of various quantities with respect to discrete transformations. With respect to rotation, chirality is invariant, while the other shown quantities transform as vectors.}
\label{fig:overview}
\end{figure}

Of particular interest are the so-called discrete symmetries with respect to spatial inversion (P), time reversal (T), charge conjugation (C), as well as combination of these. Charge conjugation is not directly relevant to our present discussion; however, the other two symmetries are important. Spatial inversion corresponds to the inversion of all three Cartesian coordinates, an operation that can be equivalently represented as mirror reflection followed by a rotation. The symmetry with respect to spatial inversion is called \textit{parity}. The symmetry with respect to time reversal describes what happens under time $t\rightarrow-t$. Figure~\ref{fig:overview} summarizes how the quantities that appear below behave under these transformations.

In the mid 20th century, it was discovered that so-called weak interactions violate both P and T symmetries, as well as the combined PT symmetry; however, the combined CPT symmetry remains an unbroken pillar of physics. We emphasize that the effects we discuss in this paper do not break any \textit{fundamental} symmetries, and specifically, neither parity nor time reversal invariance. However, apart from fundamental violation of T, there is thermodynamic irreversibility related to dissipation. This kind of apparently T-odd effects are, indeed, important for our discussion.

\subsection{Rotational invariants}
In fundamental nuclear and particle physics, a useful concept of \textit{rotational invariants} is widely used to clearly indicate the symmetry properties of a system or a process. A version of this approach can also be useful in designing and analyzing experiments as it shows the behavior of the effect under study with respect to reversals of various experimental parameters. The rotational invariants are, in fact, rank-zero toroidal multipoles discussed in~\cite{Kusunose_2024_Symmetries}.

We illustrate the concept by recalling the famous experiment by C.\,S.\,Wu and her coworkers, where parity violation in beta decay was originally discovered~\cite{Wu1957_PV}. In that experiment, $^{60}$Co nuclei were polarized and the fluxes of electrons emitted along and opposite to the spin were compared. A large difference was found. Let us designate the nuclear spin as $\vec{s}$ and the electron momentum as $\vec{p}$. The result of the experiment can be described as a nonzero average value of the quantity $\vec{s}\cdot\vec{p}$ or, in other words, a correlation between the spin of the initial nuclei and the momentum of the produced electrons. This quantity is a scalar product of two vectors (more precisely, a pseudovector and a vector), so it is automatically rotationally invariant. Let us, however, examine its properties under time (T) and spatial (P) reversals. Since both $\vec{s}$ and $\vec{p}$ change sign under T, the rotational invariant is T-even. On the other hand, since $\vec{p}$ is odd under P and $\vec{s}$ is even, the invariant is P-odd. We thus find that beta decay violates parity but preserves time-reversal invariance.

If the spins are polarized by subjecting them to a strong magnetic field $\vec{B}$, the same rotational invariant can equivalently be written as $\vec{B}\cdot\vec{p}$, from which we see that the sign of the observed effect should reverse upon the reversal of the magnetic field. 

The general approach to constructing rotational invariants is this: Write down all relevant ``letters'', i.e., quantities (that are, in general, tensors) on which the effect can be expected to depend and then form all possible independent rank-zero combinations. In general, these rotational invariants will have different properties with respect to P and T. If we are interested, for example, in a P-odd, T-even effect, the term(s) with the corresponding symmetry will report the behavior of the effect under reversals, while the other, ``wrong''-symmetry terms, particularly, the ones that are P- and T-even, will report on the background and, possibly, much bigger effects and their respective behavior with respect to reversals.

\section{Dynamic CISS}
\begin{figure}
\centering
\includegraphics[width=0.5\linewidth]{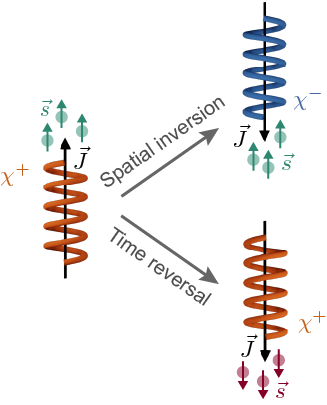}
\caption{Schematic illustration of the dynamic CISS effect: electron spins ``pushed'' through achiral medium acquire spin polarization.}
\label{fig:dynamic}
\end{figure}

The simplest dynamic CISS effect in charge transport is this: when an electric current is driven through a chiral molecule, the transmitted electrons are spin polarized as illustrated in Fig.\,\ref{fig:dynamic}. This can be represented by the rotational invariant: $\vec{s}\cdot\vec{J}$. 
Here, $\vec{s}$ is the expectation value of the electron spin and $\vec{J}$ is the electric current. This correlation is T-even (as both $\vec{s}$ and $\vec{J}$ are T-odd) and P-odd (as $\vec{s}$ is P-even and $\vec{J}$ is P-odd). The existence of such a correlation would indicate parity violation, however, it is not yet the ``full story''. Indeed, we have not yet included any mention of chirality, whereas, in the experiment, the sign of the electron polarization should reverse upon the reversal of the chirality of the molecules. Both problems are solved by writing the full correlation as 
\begin{equation}
    \textrm{Rotational invariant for dynamic CISS: } \chi\vec{s}\cdot\vec{J}\,.\label{Eq:RotInv_Transport_CISS}
\end{equation}
Here, $\chi$ is a \textit{pseudoscalar}, i.e., a scalar quantity that reverses sign upon spatial inversion, that characterizes chirality. In our example in the Introduction, this would be related to whether our randomly oriented screws are left- or right-handed.
While there is no consensus yet on an accurate \textit{measure} of chirality, the pseudoscalar $\chi$ should invert in sign between opposite enantiomers and be zero for achiral systems.

The rotational invariant \eqref{Eq:RotInv_Transport_CISS} is P- and T-even and fully describes the phenomenology of dynamic CISS, importantly, including the expected behavior under the reversals of the experimental parameters: the direction of the current and molecular chirality as illustrated in Fig.\,\ref{fig:dynamic}.

A rotational invariant tells us a lot about the behavior of the effect under various \textit{reversals} of experimental parameters; however, it does not describe the underlying physical mechanism responsible for the effect. The appearance of spin and momentum (contained in the current) strongly suggests that spin-orbit interactions play an essential role in CISS. 

\section{Quasi-static CISS}
\begin{figure}
\centering
\includegraphics[width=0.5\linewidth]{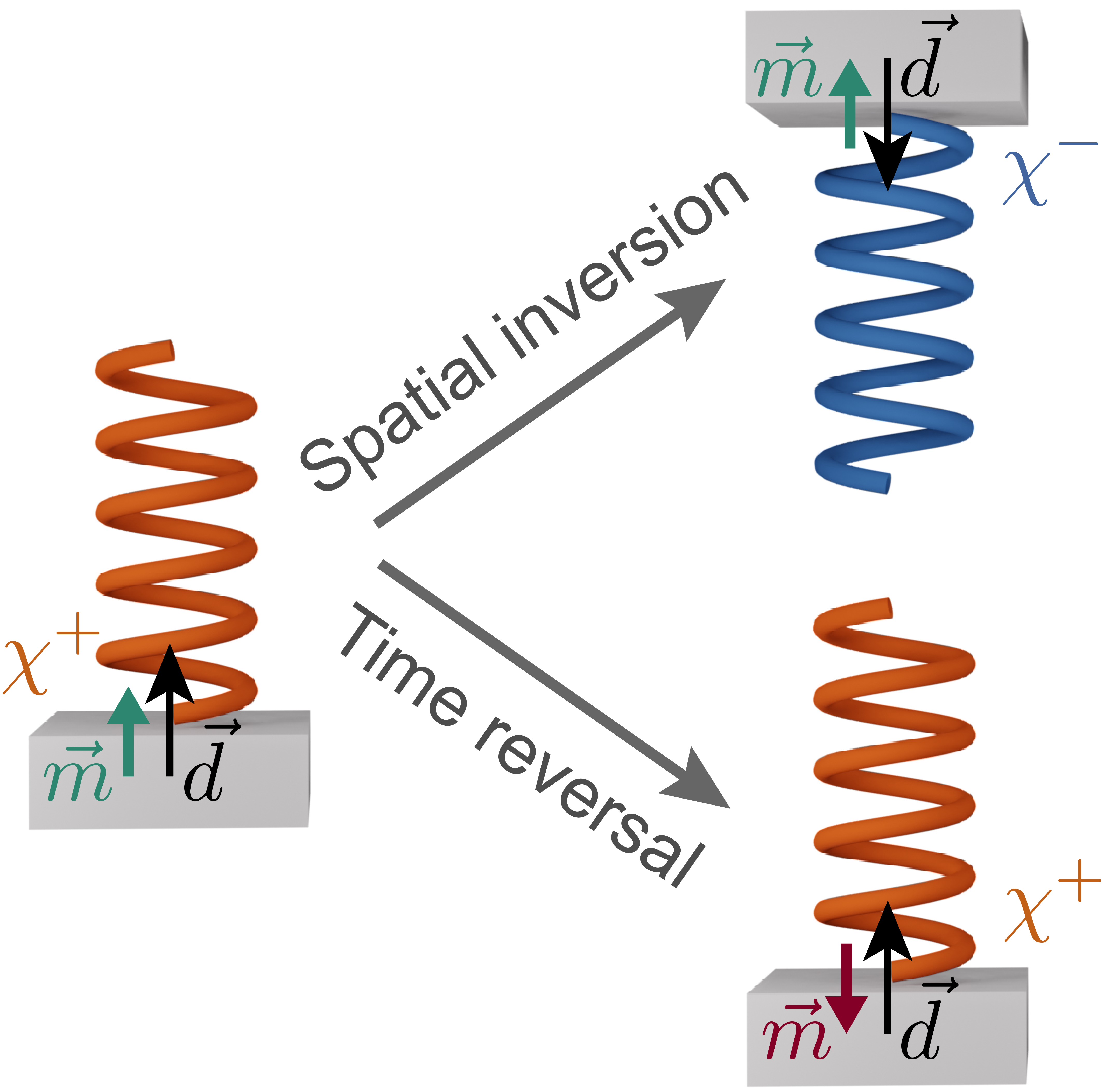}
\caption{Schematic illustration of the magneto-chiral effect: applying a chiral overlayer on top of a magnetic layer affects the magnetization of the latter. The effect is non-invariant with respect to time reversal but this is due to dissipation rather than to fundamental violation of T-invariance.}
\label{fig:static}
\end{figure}

Chirality is a T-even, P-odd pseudoscalar, while magnetization is a T-odd, P-even pseudovector. Thus, it is a priori not obvious that applying a chiral overlayer on a magnetic layer would have influence on the magnetization of the latter  as illustrated in Fig.\,\ref{fig:static}. Yet, this is exactly what has been observed in experiments~\cite{BenDor2017}.

We can try to understand this phenomenon from the point of view of symmetries. A key point is that an interface can introduce additional vectors into the problem. For example, the interaction of the chiral molecule with the magnetic material (with magnetization $\vec{m}$) can polarize the molecule, inducing an electric dipole moment $\vec{d}$. Of the quantities $\vec{m},\vec{d}$ and $\chi$, we can build a rotational invariant $\chi(\vec{d}\cdot\vec{m})$ that is P-even. The remaining issue is that this rotational invariant is also T-odd, while we expect correlations that do not involve weak interactions to be both P and T even.

The final ingredient that we need to resolve the conundrum (and restore the fundamental T-invariance) is to realize the role of \textit{dissipation}. If dissipation is involved, apparently T-odd correlations become allowed. A famous example of apparent T-violation related to dissipation is Ohm's law, $\vec{j}=\sigma\vec{E}$, where $\vec{j}$ is the current density, $\vec{E}$ is the electric field, $\sigma$ is the conductivity, and we consider current in an isotropic medium. Indeed, Ohm's law suggests a rotational invariant $\vec{j}\cdot\vec{E}$, which is T-odd since $\vec{j}$ is T-odd, while $\vec{E}$ is T-even. 
We can alleviate this apparent T-violation by defining the rotational invariant for Ohm's law as $i\vec{j}\cdot\vec{E}$.
Here the imaginary number $i$ is written to schematically indicate a sign flip upon the reversal of time. This is because such reversal interchanges the initial and final states, mathematically represented by transposition and complex conjugation. The appearance of such a coefficient in the rotational invariant is an indication that dissipation is at play in the process.

Applying this to the quasi-static CISS effect, we can write the correlation in the form 

\begin{equation}
    \textrm{Rotational invariant for quasi-static CISS: } i\chi(\vec{d}\cdot\vec{m}),\label{Eq:RotInv_Static_CISS}
\end{equation}
to emphasize the P and T invariance. Alas, the quasi-static CISS effect has nothing to do with fundamental T-violation or weak interactions.

\section{Conclusions and outlook}
We hope that in this perspective, we have been able to illustrate how fundamental concepts from the physics of elementary particles can be helpful for understanding condensed matter phenomena. 
By constructing suitable rotational invariants, we find that the CISS effects do not violate any fundamental symmetries. This approach also helps to clarify, for example, when dissipation plays an essential role in a particular effect and when it does not. Although rotational invariants are a very powerful tool to understand the underlying symmetry properties, they do not yield a quantitative prediction of the magnitude of the CISS effects.

Apart from didactical benefit, such cross-disciplinary insights and interactions between scientists working in diverse fields could lead to new areas of research, with likely practical benefits. A fitting example is the ideas of employing the CISS effect to hyperpolarize atomic nuclei \cite{Stern_2025_Conductive}. Such an approach combined with zero- to ultralow-field nuclear magnetic resonance (ZULF NMR) techniques \cite{Barskiy2025_PNMR_ZULF} could lead to truly affordable and compact NMR diagnostics tools for chemistry, biology, and medical diagnostics. Perhaps, one day you will carry it in your cell phone or watch.

\medskip
\section*{Acknowledgments}
The authors are grateful to R.\,Naaman, M.\,Kozlov, Y.\,Band, Y.\,Paltiel, and T.\,Le for critical reading of the manuscript.
This work was supported by the EU, project HEU-RIA-MUQUABIS-101070546. We acknowledge support from the Deutsche Forschungsgemeinschaft (DFG, German Research Foundation) via TRR 173 Spin+X 268565370 (Projects A09, B14), and SFB 1552/1 465145163 (Projects A01, A02, Q01), and Project ID 390831469: EXC 2118 (PRISMA++ Cluster of Excellence).

\bibliography{references}

\end{document}